\documentclass[prd,preprint,superscriptaddress,amsmath,amssymb,nofootinbib]{revtex4}
\usepackage{graphicx}
\usepackage{dcolumn}
\usepackage{bm}
\usepackage{amssymb}
\usepackage{amsmath}
\usepackage{epsfig}    
\usepackage{color}
\usepackage{slashed}
\usepackage{hhline}

\def\be{\begin{equation}}
\def\ee{\end{equation}}
\newcommand{\bea}{\begin{eqnarray}}
\newcommand{\eea}{\end{eqnarray}}
\newcommand{\nn}{\nonumber}

\newcommand{\ET}{\mbox{$\not \hspace{-0.10cm} E_T$ }}

\begin{document}

\begin{flushright}{KIAS-P18059, APCTP Pre2018 - 009} \end{flushright}

\title{A Linear seesaw model with hidden gauge symmetry}

\author{Takaaki Nomura}
\email{nomura@kias.re.kr}
\affiliation{School of Physics, KIAS, Seoul 02455, Republic of Korea}

\author{Hiroshi Okada}
\email{macokada.hiroshi@apctp.org}
\affiliation{Asia Pacific Center for Theoretical Physics, Pohang, Geyoengbuk 790-784, Republic of Korea}

\date{\today}

\begin{abstract}
 We propose a natural realization of linear seesaw model with hidden gauge symmetry
 in which $SU(2)_L$ triplet fermions, one extra Higgs singlet, doublet and quartet scalar are introduced.
 Small neutrino mass can be realized by two suppression factors that are small vacuum expectation value of quartet scalar and inverse of Dirac mass for triplet.
 After formulating neutrino mass matrix, we discuss collider phenomenology of the model focusing on signals from exotic charged particles production at the LHC.
 \end{abstract}
\maketitle

\section{Introduction}

One of the big mystery in the standard model (SM) of particle physics is the mass spectrum and flavor structure of fermions.
In particular, existence of physics beyond the SM is required from at least two non-zero neutrino masses for its generating mechanism.
Moreover, the neutrino mass indicates a hint of structure of new physics as it should explain the smallness of the mass.
Actually many mechanisms to generate neutrino mass are proposed such as canonical seesaw model~\cite{Seesaw1, Seesaw2, Seesaw3, Seesaw4}, inverse seesaw model~\cite{Mohapatra:1986bd, Wyler:1982dd}, linear seesaw model~\cite{Wyler:1982dd, Akhmedov:1995ip, Akhmedov:1995vm}, etc. 
{Note here that mass hierarchies in the neutral mass matrix are always assumed in order to get sizable neutrino mass.
 Thus appropriate explanations about these hierarchies are also one of the important tasks in our models and there exist several explanations~\cite{Das:2017ski, Wang:2015saa}.  
In light of the motivation,} 
one interesting scenario is to generate neutrino mass using the exotic fields which are large $SU(2)_L$ multiplets like quartet, quintet or septet~\cite{Kumericki:2012bh, Law:2013gma, Yu:2015pwa, Nomura:2016jnl,Nomura:2017abu, Wang:2016lve, Nomura:2018cle}, since we can suppress neutrino mass by small vacuum expectation value (VEV) of a large multiplet scalar and/or restricted structure of interactions including large multiplet fields.
Furthermore, this kind of scenario would induce interesting phenomenology at collider experiments, since a large multiplet field contain multi-charged particles such as doubly-charged scalar/fermion.

In this letter we propose a natural realization of linear seesaw model with hidden gauge symmetry in which $SU(2)_L$ triplet fermions, one extra Higgs singlet, doublet and quartet scalar are introduced.
Interestingly, tiny neutrino mass is realized by two suppression effects; inverse of Dirac mass for the triplet fermion and small VEV of the quartet scalar which is required by the constraint from $\rho$-parameter, where the quartet VEV is induced in a similar way to the Higgs triplet model~\cite{Ref:Type-II}.
We formulate neutrino mass matrix and estimate typical size of Yukawa coupling constants associated with triplet fermion and SM leptons.
Then we discuss collider phenomenology of our scenario focusing on production of exotic charged particles at the large hadron collider (LHC).
Particularly interesting signals come from Yukawa interaction associated with triplet fermion, quartet scalar and SM lepton which represent a specific signature of our model.

This letter is organized as follows.
In Sec. II, we introduce our model, and formulate Higgs sector, neutral gauge sector, neutrino sector{, and relevant interactions}.
In Sec.III, we discuss collider phenomenologies of exotic charged particles considering specific signature in our model.
 Finally we devote the summary of our results and the conclusion in Sec.IV.

\section{Model setup and Constraints}
\begin{table}[t!]
\begin{tabular}{|c||c|c|c|c|c|c|c||c|c|c|c|}\hline\hline  
& ~$Q_{L_a}$~& ~$u_{R_a}$~  & ~$d_{R_a}$~& ~$L_{L_a}$~& ~$e_{R_a}$~& ~$\Sigma_{R_a}$~& ~$\Sigma_{L_a}$~& ~$H$~& ~$H_1$~& ~$H_4$~& ~$\Phi$~\\\hline\hline 
$SU(3)_C$ & $\bm{3}$  & $\bm{3}$ & $\bm{3}$ & $\bm{1}$ & $\bm{1}$ & $\bm{1}$ & $\bm{1}$ & $\bm{1}$ & $\bm{1}$ & $\bm{1}$ & $\bm{1}$  \\\hline 
$SU(2)_L$ & $\bm{2}$  & $\bm{1}$  & $\bm{1}$  & $\bm{2}$  & $\bm{1}$  & $\bm{3}$  & $\bm{3}$ & $\bm{2}$   & $\bm{2}$ & $\bm{4}$& $\bm{1}$   \\\hline 
$U(1)_Y$   & $\frac16$ & $\frac23$ & $-\frac13$ & $-\frac12$  & $-1$ & $0$  & $0$  & $\frac12$ & $\frac12$  & $-\frac12$  & $0$\\\hline
$U(1)_{H}$   & $0$ & $0$ & $0$   & $0$  & $0$  & $1$  & $1$  & $0$ & $1$  & $1$  & $1$\\\hline
\end{tabular}
\caption{ 
Charge assignments of our fields
under $SU(3)_C\times SU(2)_L\times U(1)_Y\times U(1)_{H}$, where the lower index $a$ is the number of family that runs over 1-3.}
\label{tab:1}
\end{table}

In this section, we formulate our model introducing hidden gauge symmetry $U(1)_H$.
At first, we add three families of $SU(2)_L$ triplet right(left)-handed fermions $\Sigma_R(\Sigma_L)$ with $U(1)_H$ charge $1$; the triplet fermions can satisfy anomaly cancellation conditions since they are vector-like.
In scalar sector, we introduce three new scalar fields in addition to the SM Higgs field, which are $SU(2)_L$ doublet $H_1$, $SU(2)_L$ quartet $H_4$ and SM singlet $\varphi$ with $U(1)_H$ charge $1$.
Here we denote each VEV of the scalar fields to be $\langle H\rangle\equiv v_H/\sqrt2$, $\langle H_1\rangle\equiv v_1/\sqrt2$,  $\langle H_4\rangle\equiv v_4/\sqrt2$, and $\langle \Phi\rangle\equiv v_{\Phi}/\sqrt2$, where $H$ is expected to be the SM-like Higgs.
All the field contents and their charge assignments are summarized in Table~\ref{tab:1}.

We write the singlet and doublet scalar fields by
\begin{equation}
\Phi = \frac{1}{\sqrt{2}} (v_\Phi + \phi + i \eta_{\Phi}), \quad 
H =\left( \begin{array}{c} h^+ \\ \frac{v_H+ \tilde h+i \eta }{\sqrt2} \end{array}\right), \quad
H_1 =\left( \begin{array}{c} h_1^+ \\ \frac{v_1 +h_1+i \eta_1 }{\sqrt2} \end{array}\right).
\end{equation}
The quartet scalar $H_4$ with hypercharge $Y=-1/2$ is represented as 
\begin{equation}
H_4 = \left( \varphi^{+}_1, \varphi^{0}, \varphi^{-}_2, \varphi^{--} \right)^T, \quad {\rm or} \quad
(H_4)_{ijk}, \label{eq:quartet}
\end{equation}
where subscripts for singly charged component distinguish two independent fields, and $(H_4)_{ijk}$ is the symmetric tensor notation with $SU(2)_L$ index $\{i,j,k\}$ taking $1$ or $2$, defined by
$(H_4)_{[111]} = \varphi^{+}_1$, $(H_4)_{[112]} = \varphi^{0}/\sqrt{3}$, $(H_4)_{[122]} = \varphi^{-}_2/\sqrt{3}$ and $(H_4)_{[222]} = \varphi^{--}$; $[ijk]$ indicates {symmetric indices} under exchange among them.
Note also that neutral component is written by $\varphi^0 = (v_4 + \varphi^0_R + i \varphi^0_I)/\sqrt{2}$.
The triplet fermion with hypercharge $Y=0$ is given by 
\begin{equation}
\Sigma_{L_a(R_a)} =\left(
\begin{array}{cc}
\frac{\Sigma^0}{\sqrt2} & \Sigma^{+}\\
\Sigma'^{-} & -\frac{\Sigma^0}{\sqrt2},
\end{array}\right)_{L_a(R_a)}
\end{equation}
where two charged components are distinguished as independent fermions~\footnote{We can also write $\Sigma$ as symmetry tensor form $\Sigma_{11} = \Sigma^+$, $\Sigma_{12} = \Sigma_{21} = \Sigma^0/\sqrt{2}$ and $\Sigma'^- = \Sigma_{22}$.}.
The mass of $\Sigma$ is given by Dirac type:
\begin{equation}
M_\Sigma {\rm Tr}[\bar \Sigma \Sigma] = M_\Sigma (\bar \Sigma^+ \Sigma^+ + \bar \Sigma^0 \Sigma^0 + \bar \Sigma'^- \Sigma'^-),
\end{equation}
where we have omitted flavor index.
 Note that Majorana mass term of the triplet fermions is forbidden by $U(1)_H$ symmetry and type-III seesaw mechanism is absent in our setup.

The relevant Yukawa Lagrangian under these symmetries is given by~\footnote{Since the structure of quark sector is exactly same as the one in the SM, we neglect it hereafter.} 
\begin{align}
-{\cal L_\ell}
=  y_{\ell_{aa}} \bar L_{L_a} H e_{R_a}  +  y_{R_{ab}} [\bar L_{L_a} \tilde H_1 \Sigma_{R_b}]  +  y_{L_{ab}} [\bar L_{L_a}  H_4 \Sigma^c_{L_b}] + {\rm h.c.}, \label{Eq:yuk}
\end{align}
where $\tilde H \equiv i\sigma_2H$, and upper indices $(a,b)=1$-$3$ are the number of families, and $y_\ell$ and $M_\Sigma$ can be diagonal matrix without loss of generality due to the redefinitions of the fermions.   
Here, we explicitly write our Lagrangian in terms of each components;
\begin{align}
  y_{R_{ab}} [\bar L_{L_a}  \tilde H_1 \Sigma_{R_b}]  = &
\frac{y_{R_{ab}}}{\sqrt{2}}  \left[\bar e_{L_a} (\sqrt2  \Sigma'^-_{R_b} h_1^*+ \Sigma^0_{R_b} h_1^-) + \bar \nu_{L_a}( \Sigma^0_{R_b} h_1^*+\sqrt2 \Sigma_{R_b}^+ h_1^-) \right] , \label{eq:Yukawa1} \\
   y_{L_{ab}} [\bar L_{L_a} H_4 \Sigma_{L_b}^c]  = & \frac{ y_{L_{ab}}}{\sqrt3}
[\bar e_{L_a} (\sqrt3 \Sigma'^{-c}_{L_b} \varphi^{--} + \sqrt2 \Sigma^{0c}_{L_b} \varphi_2^- +  \Sigma^{+c}_{L_b}\varphi^{0} ) \nn \\
& \qquad + \bar \nu_{L_a}(\sqrt3 \Sigma^{+c}_{L_b} \varphi_1^+ + \sqrt2 \Sigma_{L_b}^{0c} \varphi^0 +  \Sigma'^{-c}_{L_b} \varphi_2^{-})]. \label{eq:Yukawa2}
\end{align}
From these Yukawa couplings, we obtain mass matrices defined by $m_\ell=y_\ell v_H/\sqrt2$, $m_D=y_R v_1/\sqrt{2}$, $\delta_D=y_L v_4/\sqrt3$ 
where $m_D$ and $\delta_D$ contribute to neutrino mass matrix as we discuss below.
 In our model we assign lepton number $1$ to $\Sigma_{L,R}$ and the term with $y_L$ breaks lepton number conservation. 

\subsection{Scalar sector}
The scalar potential of our model is 
\begin{align}
 {\cal V} & = - \mu_H^2 H^\dagger H + \mu_{H_1}^2 H_1^\dagger H_1 + M_4^2 H_4^\dagger H_4 - \mu_\Phi^2 \Phi^\dagger \Phi + {\cal V}_{4} +  {\cal V}_{\rm non-trivial}, \\
 {\cal V}_4 & = \lambda_H (H^\dagger H)^2 + \lambda_{H_1} (H_1^\dagger H_1)^2 + \lambda_{H_4} (H_4^\dagger H_4)^2 \nonumber \\
& + \lambda_{H H_1} (H^\dagger H)(H_1^\dagger H_1) + \lambda_{H H_4} (H^\dagger H)(H_4^\dagger H_4) + \lambda_{H_1 H_4} (H_1^\dagger H_1)(H_4^\dagger H_4),
\end{align}
where ${\cal V}_4$ indicates trivial four point interaction terms.
 The parameters in ${\cal V}_4$ are assumed to satisfy constraints from unitarity and perturbativity, and we do not discuss them in our analysis since it is not closely related to neutrino mass generation and collider physics. 
The non-trivial scalar potential terms are given by
\begin{align}
{\cal V}_{\rm non-trivial}
&= \lambda_0[ (H_4 H \tilde H_1 H)+{\rm h.c.}] + \mu_0 [(H_1^\dag H)\Phi+{\rm h.c.}], \label{Eq:pot}
\end{align}
where $SU(2)_L$ indices are implicitly contracted to be gauge invariant in the first term.
These non-trivial terms forbid dangerous massless Goldstone bosons (GBs) that would be induced from $H_{1,4}$ after symmetry breaking.
The VEVs of the scalar fields are obtained by imposing the condition $\partial \mathcal{V}/ \partial v_{H,1,4,\Phi} =0$, where we assume $M_4^2 > 0$ in the potential.
Then $v_4$ is roughly given by
\begin{equation}
v_4 \sim \frac{\lambda_0 v_1 v_H^2}{M_4^2} = \lambda_0 \left( \frac{v_1}{100 \ {\rm GeV}} \right) \left( \frac{v_H}{100 \ {\rm GeV}} \right)^2 \left( \frac{1000 \ {\rm GeV}}{M_4} \right)^2 \ {\rm GeV}.
\end{equation}
This VEV is restricted by the $\rho$-parameter which is given by 
\begin{align}
\rho=\frac{v_H^2+v_1^2 +7 v_4^2}{v_H^2+v_1^2 +  v_4^2},
\end{align}
where the experimental value is $\rho=1.0004^{+0.0003}_{-0.0004}$ at $2\sigma$ confidence level~\cite{pdg}.  
On the other hand, we also require $v\equiv v_H^2+v_1^2 + 7 v_4^2=1/(\sqrt2 G_F)\approx$ (246 GeV)$^2$.
To satisfy both of conditions, 
one finds $v_4 \lesssim$ 2.65 GeV while $v_1$ can be comparable to $v_H$. 
Remarkably, we can naturally realize $v_4 \lesssim \mathcal{O}(1)$ GeV if mass scale of scalar quartet is $\mathcal{O}(1)$ TeV or larger.
With small $v_4$, the scalar bosons from $H_4$ have approximately degenerate masses which are given by $M_4$.
In our scenario, we assume small mixing among scalar quartet, doublets and singlet, and we write mass eigenstates from $H_4$ just as 
$\{\varphi^{\pm \pm }\, \varphi^\pm_1,  \varphi^\pm_2, \varphi^0_{R}, \varphi^0_I \}$ which are approximately the same as in Eq.~(\ref{eq:quartet}).

Assuming small mixing between two Higgs doublet sector and the other scalar sector, interactions associated with two Higgs doublets are approximately the same as those in 
Type-I two Higgs doublet model (2HDM).
Then we write mass eigenstates from two Higgs doublets as $\{ h, H, A, H^\pm \}$, where $h$ is the SM-like Higgs, $H$ is heavy neutral Higgs, $A$ is CP-odd Higgs and 
$H^\pm$ is charged Higgs.
In terms of the mass eigenstates, we can write the Yukawa interactions in Eq.~(\ref{eq:Yukawa2}) such that 
\begin{align}
& \frac{y_{R_{ab}}}{\sqrt{2}}  \biggl[ \sqrt{2} \bar e_{L_a} \Sigma'^-_{R_b} (\cos \alpha h + \sin \alpha H - i \cos \beta A) + \bar e_{L_a} \Sigma^0_{R_b} \cos \beta H^- \nonumber \\
& \qquad \quad + \bar \nu_{L_a} \Sigma^0_{R_b} (\cos \alpha h + \sin \alpha H - i \cos \beta A) + \sqrt{2} \bar \nu_{L_a} \Sigma^+_{R_b} \cos \beta H^- \biggr] , \label{eq:Yukawa3} 
\end{align}
where 
$\cos \alpha (\sin \alpha)$ correspond to mixing among neutral scalars in two Higgs doublet, and $\tan \beta = v_1/v_H$.

After $U(1)_H$ symmetry breaking, CP-odd component of singlet scalar $\Phi$ is absorbed by $Z'$ boson as Nambu-Goldstone boson (NGB) while CP-even component is 
physically neutral scalar boson. Under small mixing assumption, this CP-even scalar boson does not provide any interesting phenomenologies and we will not discuss it hereafter. 

\subsection{$Z'$ boson from $U(1)_H$}

In this model, we have massive $Z'$ boson from spontaneous breaking of $U(1)_H$ gauge symmetry.
Here we assume $Z'$ mass is mostly induced by the VEV of singlet scalar $\Phi$ such that
\begin{equation}
m_{Z'} \simeq g_H v_\Phi,
\end{equation}
where $g_H$ is the gauge coupling constant associated with $U(1)_H$.
Since SM particles are not charged under the $U(1)_H$, $Z'$ is hidden gauge boson and it is difficult to directly produce it at collider experiments.
Thus we will not discuss $Z'$ boson physics further in this paper.
 
\subsection{Neutrino sector}
After the spontaneous symmetry breaking, neutral fermion mass matrix with 9$\times$9 is given by
\begin{align}
M_N
&=
\left[\begin{array}{ccc}
0 & m_D & \delta_D  \\ 
m_D^T & 0 & M_\Sigma^T \\ 
\delta_D^T  & M_\Sigma & 0 \\ 
\end{array}\right].
\end{align}
Then the active neutrino mass matrix can approximately be found as
\begin{align}
m_\nu\approx -\delta_D (M_\Sigma^T)^{-1} m_D^T -m_D M_\Sigma^{-1} \delta_D^T,
\end{align}
where $\delta_D<< M_{\Sigma}$ is expected.
Let us estimate the neutrino mass order.
If $m_D\approx {\cal O}(0.01)$ GeV, $\delta_D/M_\Sigma \approx {\cal O}(10^{-8})$ is expected to find the sizable neutrino masses; $m_\nu \sim 10^{-10}$ GeV.
Moreover, in terms of Yukawa coupling constant and VEVs, we can write
\begin{equation}
m_\nu \sim y_L y_R \frac{v_1 v_4}{M_\Sigma}.
\end{equation}
Taking $M_\Sigma = 1000$ GeV, $v_4 = 1$ GeV and $v_1 \lesssim 100$ GeV, we can realize $m_\nu \lesssim 10^{-10}$ GeV with $y_L \sim y_R \lesssim 10^{-4}$
which is similar magnitude to those in generating SM charged leptons.

The neutrino mass matrix is diagonalized by unitary matrix $U_{MNS}$; $D_\nu= U_{MNS}^T m_\nu U_{MNS}$, where $D_\nu\equiv {\rm diag}(m_1,m_2,m_3)$. 
One of the elegant ways to reproduce the current neutrino oscillation data~\cite{pdg} is to apply the Casas-Ibarra parametrization~\cite{Casas:2001sr},
and find the following relation
\begin{align}
m_D =-\frac12 \delta_D (M^T_\Sigma)^{-1} (U_{MNS}^* \sqrt{D_\nu} U^\dag_{MNS}+A).
\end{align}
Here $A$ is an arbitrary 3 by 3 anti-symmetric matrix with complex value; $A+A^T=0$.
Note here that all the components of $m_D$ should not exceed {100} GeV, once perturbative limit of $y_R$ is taken to be 1.

\noindent \underline{\it Non-unitarity}:
{Constraint of non-unitarity should always be taken into account in case of larger neutral mass matrix whose components are greater than three by three, since experimental neutrino oscillation results suggest nearly unitary.
In case of the linear seesaw, when non-unitarity matrix $U'_{MNS}$ is defined, one can typically parametrize it by the following form:}
\begin{align}
U'_{MNS}\equiv \left(1-\frac12 FF^\dag\right) U_{MNS},
\end{align}
where $F\equiv  (M_{\Sigma}^T)^{-1} m_D$ is a hermitian matrix, and $U'_{MNS}$ represents the deviation from the unitarity. 
{Considering several experimental bounds~\cite{Fernandez-Martinez:2016lgt},
one finds the following constraints~\cite{Agostinho:2017wfs}:}
\begin{align}
|FF^\dag|\le  
\left[\begin{array}{ccc} 
2.5\times 10^{-3} & 2.4\times 10^{-5}  & 2.7\times 10^{-3}  \\
2.4\times 10^{-5}  & 4.0\times 10^{-4}  & 1.2\times 10^{-3}  \\
2.7\times 10^{-3}  & 1.2\times 10^{-3}  & 5.6\times 10^{-3} \\
 \end{array}\right].
\end{align} 
Here, we show a benchmark point to satisfy the neutrino oscillation data~\cite{Forero:2014bxa}, non-unitarity constraints, and perturbativity $y_R\lesssim1$, within our parameter choices.
Fixing the following values $(v_1,v_4)=(1,100)$ GeV, the benchmark points are given by:
\begin{align}
& (A_{12},A_{13},A_{23})=(-0.0306,-8.34\times10^{-5},5.93\times10^{-4})\ {\rm GeV} ,\\
&m_D  \approx
\left[\begin{array}{ccc} 
-1.6+0.019i & -478-0.007i & -18-0.01i\\
15.8-0.17i  & 9654+0.27i & -30.1+0.27i  \\
-56+0.60i  & -27830-0.75i  & -4.26-0.77i\\
 \end{array}\right]\times 10^{-10}\ {\rm GeV},\\
 &M_\Sigma  \approx
\left[\begin{array}{ccc} 
1.48 &1.50 & 1.95\\
1.13  & 1.44 & 1.65  \\
1.04  & 1.46  & 1.37\\
 \end{array}\right]\times 10^{3}\ {\rm GeV},\
 \delta_D  \approx
\left[\begin{array}{ccc} 
0.367 &1.70 & 2.57\\
5.51  & 0.50 & 12.5  \\
18.04  & 32.8  & 0.83\\
 \end{array}\right]\times 10^{-3}\ {\rm GeV},
\end{align}
where we assumed real elements of $A$ and the normal neutrino mass ordering with vanishing neutrino mass for the lightest neutrino, for simplicity.
Even though one analyzes it with a general framework, reproducing neutrino oscillation data with these constraints can easily be achieved due to enough input parameters.

\subsection{Lepton flavor violations (LFVs) and charged-lepton mass matrix}
\label{lfv-lu}
Since $y_R$ is expected to be small from the previous discussion, 
we focus on the Yukawa term $y_L$ that gives rise to $\mu\to e\gamma$ processes
at one-loop level, which is the most stringent constraint from the MEG experiment~\cite{TheMEG:2016wtm}; therefore, its branching ratio is given by $B(\mu\rightarrow e\gamma) \leq4.2\times10^{-13}$.
While our branching ratio is given by
\begin{align}
B(\mu\to e \gamma)
\approx 
\frac{48\pi^3 \alpha_{\rm em}}{{G_{\rm F}^2 } } 
\left|\sum_{\alpha=1-3} \frac{y_{L_{1\alpha}} y^\dag_{L_{\alpha 2}} }{(4\pi)^2} 
\left[2 F(\Sigma^-_\alpha , \varphi^{--}) + F(\Sigma^0_\alpha , \varphi^{-}) \right]\right|^2
,
\end{align}
where we assume all the masses in the components of $\Sigma$ and $H_4$ to be degenerate,
$G_{\rm F}\approx1.166\times 10^{-5}$ GeV$^{-2}$ is the Fermi constant, 
$\alpha_{\rm em}(m_Z)\approx {1/128.9}$ is the 
fine-structure constant~\cite{pdg}, and
\begin{align}
F(m_1,m_2)\approx \frac{m_1^6+3 m_1^4 m_2^2-6 m_1^2 m_2^4+m_2^6+ 12m_1^4 m_2^2
\ln\left[\frac{m_2}{m_1}\right]}
{12(m_1^2-m_2^2)^4}.
\label{eq:lfv-lp}
\end{align} 
Comparing our branching ratio with the experimental one, 
one finds the following bounds on Yukawa couplings:
\begin{align}
\sum_{\alpha=1-3} y_{L_{1\alpha}} y^\dag_{L_{\alpha 2}}\lesssim 2.02\times10^{-3}.
\end{align}
where we fix $M_\Sigma=600$ GeV and $m_\varphi=$1000 GeV.

\noindent \underline{\it Charged-lepton mass matrix}:

Next, we discuss the charged-lepton mixing that is also restricted by the current experimental data.
Similar to the case of LFVs, we neglect the contribution to $y_R$ because it is sufficiently small.
Furthermore, we assume the mass matrices $m_D$ and $M_\Sigma$ to be diagonal for simplicity.
Then, one finds the charged-lepton fermion mass matrix as
\begin{align}
& \left( \begin{array}{c} \bar e_L^a \\ \bar \Sigma_L^a \end{array} \right)^T
M_E
\left( \begin{array}{c}  e_R^a \\  \Sigma_R^a \end{array} \right)
=
\left( \begin{array}{c} \bar e_L^a \\ \bar \Sigma_L^a \end{array} \right)^T
\left[\begin{array}{cc}
 m_{\ell_a} & \delta_{D_a}/\sqrt2  \\ 
0  & M_{\Sigma_a}  \\ 
\end{array}\right]
\left( \begin{array}{c}  e_R^a \\  \Sigma_R^a \end{array} \right),\\
& M_E M_E^\dag
=
\left[\begin{array}{cc}
 m_{\ell_a}^2 & \delta_{D_a} M_{\Sigma_a}/\sqrt2  \\ 
\delta_{D_a} M_{\Sigma_a}/\sqrt2 & M_{\Sigma_a}^2  \\ 
\end{array}\right], 
\end{align}
The mass matrix is diagonalized by transformation $(e_{L(R)}, E_{L(R)}) \to V_{L(R)}^\dag (e_{L(R)}, E_{L(R)})$.
Thus one obtains diagonalization matrices $V_L$ which diagonalizes $M_E M_E^\dag$ as $V_L M_EM_E^\dag V_L^\dag\sim{\rm diag}(m^2_\ell, M^2_\Sigma)$ such that 
\begin{align}
 V_L = \left( \begin{array}{cc} \cos \theta & - \sin \theta \\ \sin \theta & \cos \theta  \end{array} \right), \quad
 \tan 2 \theta \simeq \frac{\sqrt2 \delta_D}{M_\Sigma}, 
\end{align}
where the current experimental data at LHC and LEP suggests that the lower bound on the heavier is about $M_\Sigma\sim$100 GeV~\cite{pdg}, while $\delta_D\sim{\cal O}(10^{-10})\sim{\cal O}(10^{-6})$. It implies that $\theta\approx {\cal O}(10^{-12})\sim{\cal O}(10^{-8})$ which is negligibly small.

\section{Collider phenomenology of the model}

In this section, we discuss production of exotic particles in the model at the LHC.
Signals of our exotic particles are explored by estimating production cross section and formulating branching ratios.
In particular we focus on charged particles in quartet scalar and triplet fermions since they induce specific signature of the model. 

\subsection{Production cross sections}
The components of quartet scalar and triplet fermions can be produced by electroweak interaction at the LHC.
For quartet scalar $H_4$, gauge interactions are derived from kinetic term:
\begin{align}
|D_\mu H_4|^2 {\supset} \sum_{m= -\frac32,-\frac12,\frac12,\frac32} \biggl| & \left[ \partial_\mu - i \left(-\frac12+m \right) e A_\mu - i \frac{g}{c_W} \left(m - \left( -\frac12+m \right) s_W^2 \right) Z_\mu \right] (H_4)_{m} \nonumber \\
& + \frac{ig}{\sqrt{2}} \sqrt{ \left(\frac32 + m \right) \left(\frac52 -m \right) } W^+_\mu (H_4)_{m-1} \nonumber \\
& + \frac{ig}{\sqrt{2}} \sqrt{ \left(\frac32 - m \right) \left(\frac52 +m \right) } W^-_\mu (H_4)_{m+1} \biggr|^2,
\end{align}
where $g$ is the gauge coupling for $SU(2)_L$, $e$ is electromagnetic coupling constant, $s_W(c_W) = \sin \theta_W (\cos \theta_W)$ with the Weinberg angle $\theta_W$, and 
$(H_4)_m$ indicates the component of $H_4$ which has the eigenvalue of diagonal $SU(2)_L$ generator $T_3$ given by $m$; $\{(H_4)_{3/2}, (H_4)_{1/2}, (H_4)_{-1/2}, (H_4)_{-3/2} \} = \{\varphi_1^+, \varphi^0, \varphi_2^-, \varphi^{--} \}$.
For triplet fermion, we explicitly write gauge interactions such that 
\begin{align}
\mathcal{L} =& g W_\mu^+ (\bar \Sigma^+ \gamma^\mu \Sigma^0 + \bar \Sigma^0 \gamma^\mu \Sigma'^-) + g W_\mu^- (\bar \Sigma^0 \gamma^\mu \Sigma^+ + \bar \Sigma'^- \gamma^\mu \Sigma^0) \nonumber \\
& + g c_W Z_\mu (\bar \Sigma^+ \gamma^\mu \Sigma^+ - \bar \Sigma'^- \gamma^\mu \Sigma'^-) + e A_\mu (\bar \Sigma^+ \gamma^\mu \Sigma^+ - \bar \Sigma'^- \gamma^\mu \Sigma'^-).
\end{align}

\begin{figure}[tb]\begin{center}
\includegraphics[width=70mm]{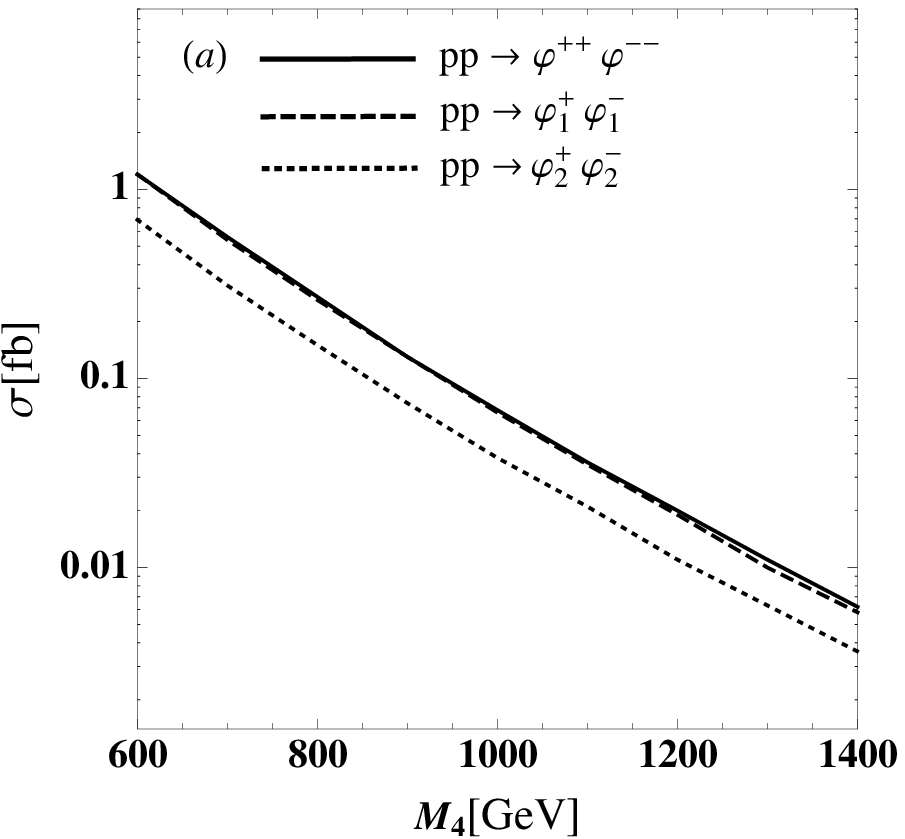}
\includegraphics[width=70mm]{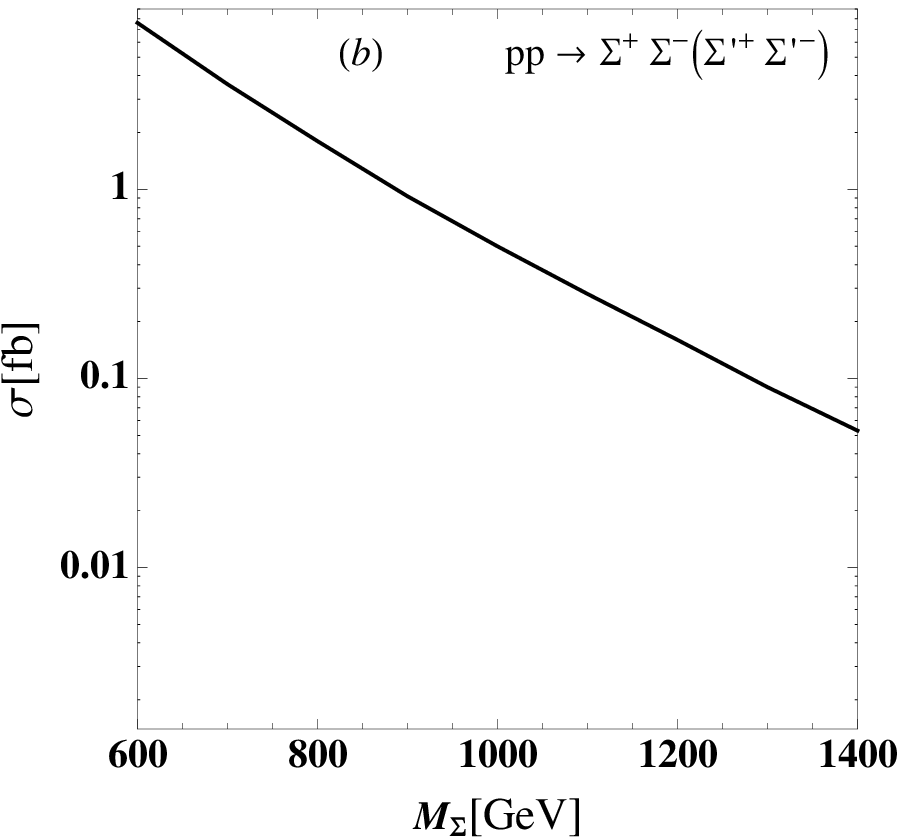}
\caption{(a)The cross section for pair production process $p p \to Z/\gamma \to \varphi^{++} \varphi^{--}$ and $pp \to Z/\gamma \to \varphi^+_{1,2} \varphi^-_{1,2}$ as a function of quartet mass $M_4$. (b)The cross section for pair production process $p p \to Z/\gamma \to  \Sigma^{+} \Sigma^{-} ( \Sigma'^+ \Sigma'^-)$ as a function of triplet fermion mass $M_\Sigma$.
}   \label{fig:CX}\end{center}\end{figure}

 The exotic charged particles can be produced via interactions with the SM gauge bosons. 
The production cross sections are estimated using CalcHEP~\cite{Belyaev:2012qa} with the CTEQ6 parton distribution functions (PDFs)~\cite{Nadolsky:2008zw}.
In Fig.~\ref{fig:CX} we show cross sections for pair production of exotic charged particles at the LHC 13 TeV.
We find that production cross section for $\Sigma^\pm (\Sigma'^\pm)$ pair production is larger than those for charged scalars in quartet when the mass scale is same.
For $\mathcal{O}(1)$ TeV mass of exotic fermions, we can obtain production cross section $\sim 1$ fb which can give sizable number of event at the LHC.

\subsection{Decay branching ratios of exotic particles}

Here, we consider decay processes of exotic charged particles and estimate their branching ratios (BRs).

Firstly, we consider decay of charged scalar bosons from quartet.
Partial decay width for the processes including $\Sigma$s in final state are given by
\begin{align}
& \Gamma_{\varphi^{++} \to \ell^+ \Sigma'^+} \simeq \Gamma_{\varphi^{+}_1 \to \nu \Sigma^+} \simeq \frac{y_L^2}{16 \pi} M_4 \left( 1 - \frac{m_\Sigma^2}{M_4^2} \right)^{2} \\
& 2 \Gamma_{\varphi^{+}_2 \to \ell^+ \Sigma^0} \simeq \Gamma_{\varphi^{+}_2 \to \nu \Sigma'^+} =  \frac{y_L^2}{24 \pi} M_4 \left( 1 - \frac{m_\Sigma^2}{M_4^2} \right)^{2}, 
\end{align}
where the masses of SM leptons are ignored and we have omitted flavor index for Yukawa coupling constant $y_L$.
Scalar bosons in quartet can also decay into two SM gauge bosons through gauge interactions
\begin{align}
|D_\mu H_4|^2 \supset & \sqrt{\frac{3}{2}} v_4 W^+ W^+ \varphi^{--} + 
\frac{g^2 v_4}{c_W} \left[ s_W^2 Z_\mu W^{+ \mu} \varphi^-_2 + \frac{\sqrt{3}}{2} c_W^2    Z_\mu W^{+ \mu } \varphi^-_1 \right] \nonumber \\
& + e g v_4 \left[ A_\mu W^{+ \mu} \varphi_2^- - \frac{\sqrt3}{2} A_\mu W^{+ \mu} \varphi_1^- \right] + c.c. \, .
\end{align} 
Then we derive partial decay widths for two gauge boson final states such that
\begin{align}
& \Gamma_{\varphi^{++} \to W^+ W^+} = \frac{3 g^4 }{32 \pi} \frac{v_4^2}{M_4} \lambda^{\frac12}(M_4; m_W, m_W) \left[ 2 + \frac{M_4^4}{m_W^4} \left(1 - \frac{2 m_W^2}{M_4^2} \right)^2 \right], \\
& \Gamma_{\varphi_{1}^+ \to W^+ Z} = \frac{3 c_W^2 g^4 }{64 \pi} \frac{v_4^2}{M_4} \lambda^{\frac12}(M_4; m_W, m_Z) \left[ 2 + \frac{M_4^4}{m_W^2 m_Z^2} \left(1 - \frac{ m_W^2}{M_4^2} - \frac{ m_Z^2}{M_4^2 }\right)^2 \right], \\
& \Gamma_{\varphi_{2}^+ \to W^+ Z} = \frac{ s_W^4 g^4 }{16 c_W^2 \pi} \frac{v_4^2}{M_4} \lambda^{\frac12}(M_4; m_W, m_Z) \left[ 2 + \frac{M_4^4}{m_W^2 m_Z^2} \left(1 - \frac{ m_W^2}{M_4^2} - \frac{ m_Z^2}{M_4^2 }\right)^2 \right], \\
& \Gamma_{\varphi_{1}^+ \to W^+ \gamma} = \frac{3 g^2 e^2 }{64 \pi} \frac{v_4^2}{M_4} \left( 1 - \frac{m_W^2}{M_4^2} \right), \\
& \Gamma_{\varphi_{2}^+ \to W^+ \gamma} = \frac{ g^2 e^2 }{16 \pi} \frac{v_4^2}{M_4} \left( 1 - \frac{m_W^2}{M_4^2} \right),
\end{align}
where the factor $\lambda(m_1; m_2, m_3)$ is defined as 
\begin{equation}
 \lambda(m_1; m_2, m_3) \equiv 1 + \frac{m_2^4}{m_1^4} + \frac{m_3^4}{m_1^4} - \frac{2 m_2^2}{m_1^2} - \frac{2 m_3^2}{m_1^2} - \frac{2 m_2^2 m_3^2}{m_1^4}.
\end{equation}
%
\begin{figure}[tb]\begin{center}
\includegraphics[width=70mm]{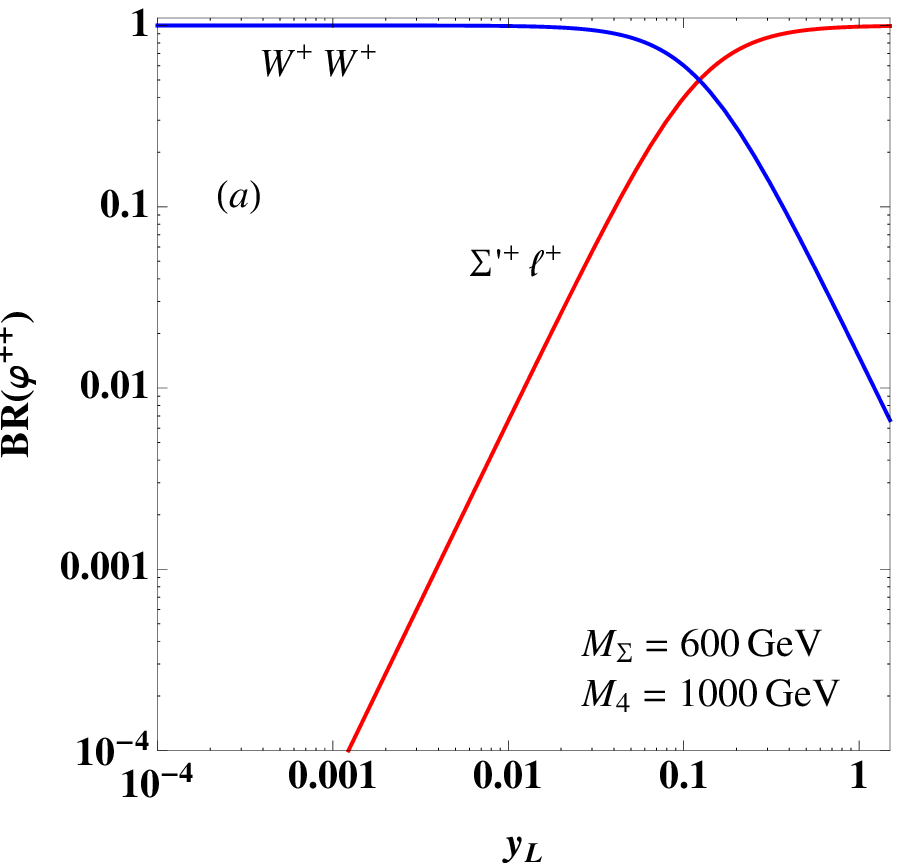}
\includegraphics[width=70mm]{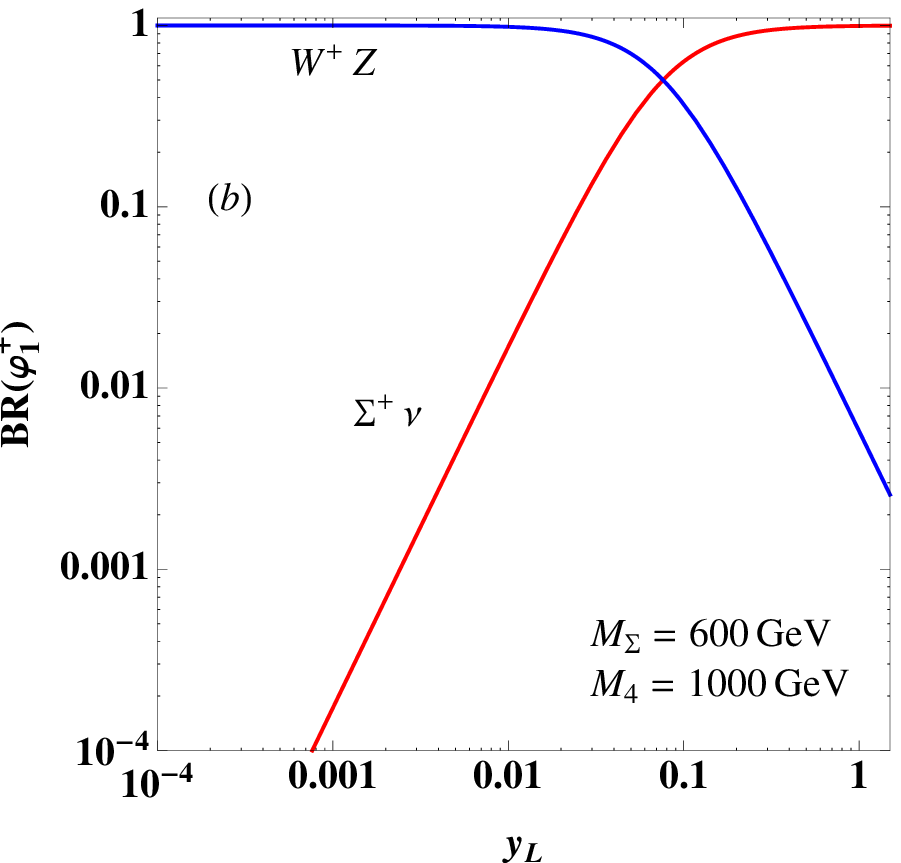}
\caption{(a) Branching ratio for decay of $\varphi^{++}$ in quartet scalar. (b) Branching ratio for decay of $\varphi_1^+$ in quartet scalar. 
They are given as a function of Yukawa coupling $y_L$ assuming one element is dominant and some parameters are fixed as indicated in the plots.
}   \label{fig:BR}\end{center}\end{figure}
%
In Fig.~\ref{fig:BR}, we show the BRs for $\varphi^{++}$ and $\varphi^{+}_1$ as a function of the Yukawa coupling $y_L$ where we assume only one element of $y_{L_{ab}}$ dominates for simplicity, 
and we fixed some parameters such as $v_4 = 1$ GeV, $M_\Sigma = 600$ GeV and $M_4 = 1000$ GeV. 
We find that BR for two massive gauge boson mode is dominant if the Yukawa coupling is $y_L \lesssim 0.1$
{, where $W^+ \gamma$ mode in decay of $\varphi^+_{1}$ is negligible since it is found to be always less than $\sim 10^{-5}$.
}
The BRs for $\varphi_2^+$ and $\varphi^0$ have similar behavior where $\varphi^0_R$ can decay into $ZZ$ while $\varphi^0_I$ cannot decay into two gauge boson.

The exotic fermions decay into SM lepton and scalar bosons through Yukawa interaction in Eqs.~(\ref{eq:Yukawa2}) and (\ref{eq:Yukawa3}). 
For $M_4 > M_\Sigma$, the dominant decay modes include only scalar boson from doublet $H_1$ such as $\Sigma^+ \to H^+ \nu$ etc.
Here we choose $M_\Sigma$ is larger than scalar boson masses from doublet fields.
On the other hand $\Sigma$s can decay into both scalar bosons from $H_4$ and from doublets for $M_4 < M_\Sigma$ where the BRs are determined by relative values of $y_L$ and $y_R$ Yukawa coupling constants.
In general, decay widths of $\Sigma$s are given by
\begin{equation}
\Gamma_{\Sigma \to \Phi \ell (\nu)} \simeq \frac{C_{\Sigma \Phi \ell(\nu)}^2}{16 \pi} M_\Sigma \left( 1 - \frac{M_4^2}{M_\Sigma^2} \right)^2, \label{eq:width-sigma}
\end{equation} 
where $\Phi = \{ \varphi^{\pm \pm}, \varphi^{\pm}_{1,2}, \varphi^0_{R,I}, h, H, A, H^\pm \}$ and $\Sigma = \{\Sigma^\pm, \Sigma'^\pm, \Sigma^0 \}$ with possible charge combination in final states, 
and $C_{\Sigma \Phi \ell(\nu)}$ denotes the coupling of an interaction $\Sigma$-$\Phi$-$\ell(\nu)$ in Eqs.~(\ref{eq:Yukawa2}) and (\ref{eq:Yukawa3}).
We note that charged component $\Sigma^\pm(\Sigma'^\pm)$ can decay into $\Sigma^0 \pi^\pm$, where $\pi^\pm$ is induced from off-shell $W$ boson since 
mass difference between charged and neutral component is induced at one loop level~\cite{Cirelli:2005uq}. The mass difference is obtained as $\Delta M \sim 166$ MeV for triplet fermion 
and partial decay width is estimated to be $\Gamma_\pi \sim 3 \times 10^{-15}$ GeV. 
This width is much smaller than those in Eq.~(\ref{eq:width-sigma}) as we obtain $\Gamma_{\Sigma \to \Phi \ell (\nu)} \sim 10^{-7} $ GeV with $C_{\Sigma \Phi \ell(\nu)} = 10^{-4}$, $M_\Sigma = 1000$ GeV and $M_4 = 500$ GeV.  We can thus neglect the decay mode with pion in our analysis.

\subsection{Signals at the LHC}

Here we discuss signals of our model at the LHC focusing on charged particles in quartet scalar and triplet fermion.
The charged scalar boson from $H_4$ dominantly decays into two SM gauge bosons 
since Yukawa coupling constant $y_L$ tends to be much smaller than $\sim 0.1$ to obtain active neutrino mass consistent with observations.
Thus signal processes will be
\begin{align}
pp \to \varphi^{++} \varphi^{--} \to W^+ W^+ W^- W^-, \quad pp \to \varphi^{+}_{1,2} \varphi^{-}_{1,2} \to W^+ W^- Z Z.
\end{align}
Then $W^\pm$ and $Z$ bosons further decay into either jets or leptons.
For such a signal, detailed discussions are found in, for example, refs.~\cite{delAguila:2013yaa,delAguila:2013mia, Nomura:2017abu,Chala:2018ari}.
Then we focus on signals from exotic charged fermion production hereafter.

We consider two cases of mass relation in considering the charged fermion $\Sigma^\pm(\Sigma'^\pm)$; (A) $M_4 > M_\Sigma$, (B) $M_4 < M_\Sigma$.
In case (A), $\Sigma^\pm(\Sigma'^\pm)$ always decay into SM lepton and  scalar boson associated with Higgs doublet. 
On the other hand, in case (B), the charged fermions can decay also into SM lepton and scalar boson associated with quartet scalar 
in addition to the mode in case (A).
The signals in case (A) are then obtained as decay modes of $\Sigma^\pm(\Sigma'^\pm)$ which are $\ell^\pm \{H,h,A\}$ and $\nu H^\pm$; 
then scalar bosons further decay into SM fermions or gauge bosons where the corresponding BRs are the same situation in the Type-I 2HDM.
Here we focus on the signals of SM lepton with a component of $H_4$ in case (B), since they are specific signature of our model.
The possible final states from produced $\Sigma^+ \Sigma^-$ and $\Sigma'^+ \Sigma'^-$ pairs are summarized in Table.~\ref{tab:2} with their fractions obtained from product of BRs for each particle.
Here we assume $M_4 < M_\Sigma$, {$ y_R \ll y_L \lesssim 0.1$}
and $v_4 = 1$ GeV and we do not distinguish neutrino and anti-neutrino for simplicity; 
we include decay of components of $H_4$ into SM gauge bosons.
\begin{table}[t!]
\begin{tabular}{|c||cccccc|}\hline\hline  
From $\Sigma^+ \Sigma^-$ & ~$\ell^+ \ell^- Z Z Z Z$~ & ~$\ell^+ \ell^- Z Z \varphi^0_I$~ & ~$\ell^\pm \nu Z Z Z W^\mp$~ & ~$\ell^+ \ell^- \varphi_I^0 \varphi_I^0$~ &  ~$\ell^\pm \nu Z W^\mp \varphi_I^0$~ 
& ~$\nu \nu W^+ W^- Z Z$~\\\hline\hline 
fractions & 0.016 & 0.032 & 0.094 & 0.016 & 0.094 & 0.56 \\ \hline \hline
From $\Sigma'^+ \Sigma'^-$ & \multicolumn{2}{c}{ $\ell^+ \ell^- W^+ W^+ W^- W^-$} & \multicolumn{2}{c}{$\ell^\pm \nu Z W^\mp W^\mp W^\pm $} & ~$\nu \nu Z Z W^+ W^-$~ &  \\\hline\hline 
fractions & \multicolumn{2}{c}{0.25} & \multicolumn{2}{c}{0.25} & 0.25 & \\ \hline 
\end{tabular}
\caption{ The possible final states from $\Sigma^+ \Sigma^-$ and $\Sigma'^+ \Sigma'^-$ with their fractions given by product of BRs under the assumption of $M_4 < M_\Sigma$, $y_L \gg y_R$, $y_L \lesssim 0.1$ and $v_4 = 1$ GeV.}
\label{tab:2}
\end{table}
Note that we remain $\varphi_I$ in the table since it cannot decay into SM gauge bosons but decay into SM fermions via mixing with the CP-odd component in Higgs doublet sector.

For $\Sigma^+ \Sigma^-$ production, we obtain the largest number of events from $\nu \nu W^+ W^- Z Z$ final state. 
When $W^\pm$ and $Z$ bosons from one of $\Sigma^\pm$ decay into leptons and the other gauge bosons decay into jets signal event up to the detector level is given by
\begin{equation}
pp \to \Sigma^+ \Sigma^- \to \varphi^+_1 \bar \nu  \varphi_1^- \nu  \to   W^+ Z \bar \nu W^- Z \nu  \to \ell^\pm \ell^+ \ell^- 4j \ET,
 \end{equation}
where $j$ indicates jet and $\ET$ is missing transverse energy. 
 Thus our signal is multi-lepton with jets and missing transverse energy. 
For $M_\Sigma = 600$ GeV, the number of events without kinematical cut {can} be estimated as $\sim 8$, taking integrated luminosity as $300$ fb$^{-1}$. 
Although the number of event is not large, we would find the signal since the number of SM background (BG) events is expected to be small.
In addition, we can partially reconstruct mass of $\Sigma^\pm$ from $\ell^\pm \ell^+ \ell^- \ET$.  
In the high-luminosity LHC (HL-LHC) experiments, we can obtain more events and more parameter region will be explored.

For $\Sigma'^+ \Sigma'^-$ production, the most clear signal would come from the final state $\ell^+ \ell^- W^+ W^+ W^- W^-$. 
When $W^\pm$ bosons from one $\Sigma'^\pm$ decay into leptons and the other gauge bosons decay into jets signal event up to the detector level is given by
\begin{equation}
pp \to \Sigma'^+ \Sigma'^- \to \varphi^{++} \ell^- \varphi^{--}   \ell^+ \to \ell^- W^+ W^+ \ell^+  W^- W^- \to \ell^+ \ell^- \ell^\pm  \ell^\pm 4j \ET.
 \end{equation} 
For $M_\Sigma = 600$ GeV, we obtain the production cross section $\sigma(pp \to \Sigma'^+ \Sigma'^-) \simeq 7.6$ fb as shown in Fig.~\ref{fig:CX}.
The products of production cross section and BRs is then $\sigma(pp \to \Sigma'^+ \Sigma'^-) BR(\Sigma'^\pm \to W^{\pm} W^{\pm} \ell^\mp)^2 BR(W^\pm \to \ell^\pm \nu)^2 BR(W^\pm \to j j) \simeq 0.04$ fb where we applied $BR(\Sigma'^\pm \to W^{\pm} W^{\pm} \ell^\mp) \simeq 0.5$, $BR(W^\pm \to \ell^\pm \nu) \simeq 0.22$, and $BR(W^\pm \to j j) \simeq 0.67$. 
Thus number of events without kinematical cut {can}  be estimated as $\sim 12$, taking integrated luminosity as $300$ fb$^{-1}$. 
This cross section would be around same order as background (BG) process; for example we obtain cross section for $pp \to W^+ W^- \ell^+ \ell^- \to \ell^+ \ell^+ \ell^- \ell^- \nu \nu$ as 0.18 fb in the SM estimated by {\tt MADGRAPH5}~\cite{Alwall:2014hca}.
This signature is clearer than the previous case, since the number of events is larger and final state includes three same sign leptons.
In this case, we can partially reconstruct mass of $\Sigma'^\pm$ from $\ell^\mp \ell^\pm \ell^\pm \ET$.
However it is not trivial to select three charged leptons to reconstruct the mass from four charged leptons in the final state and we need to perform detailed analysis. 
In addition, we need to impose appropriate tagging and kinematical cuts to reduce BG events for getting sufficient significance; for example jet tagging will be useful to reduce $W^+W^- \ell^+ \ell^-$ BG
and cuts regarding angles among charged leptons can be used to choose charged leptons as decay products of one $\Sigma'^\pm$. 
Furthermore detector level simulation is required to take into account detector efficiency in order to obtain realistic number of events at the experiments.
The detailed simulation study including BG events and kinematical cuts are beyond our scope of this paper, since the final states contain many particles and 
the analysis will be very complicated.

\section{Summary and Discussions}

We have constructed a model with hidden $U(1)$ gauge symmetry which can naturally realize linear seesaw mechanism by introducing $SU(2)_L$ triplet Dirac fermion and quartet scalar fields. 
Then an induced active neutrino mass is suppressed by two factors; small VEV of quartet scalar and inverse of TeV scale Dirac mass for triplet fermion, where small quartet VEV is also required by the $\rho$-parameter constraint.
Furthermore, small VEV of the quartet is naturally realized by mechanism similar to Higgs triplet model.

We have formulated active neutrino mass matrix with linear seesaw mechanism which is given by Yukawa coupling constants associated with interactions among triplet fermion, quartet scalar and SM leptons. 
Then typical size of Yukawa coupling constants have been estimated to realize the small neutrino mass of $\mathcal{O}(0.1)$ eV.
We have found that the size of coupling can be $\mathcal{O}(10^{-5})$-$\mathcal{O}(10^{-4})$ which is similar to those in generating SM charged leptons.

Then we have discussed collider phenomenology of the model focusing on production of exotic charged particles.
Specific signatures of our model  are obtained, when an exotic charged fermion decays into SM lepton and a scalar boson from quartet which dominantly decay into SM gauge bosons.
Then our signals are multi-leptons with jets and missing transverse energy.
We have found that the number of signal events is $\mathcal{O}(10)$ at the LHC 13 TeV with integrated luminosity of 300 fb$^{-1}$, when triplet fermion mass is $\sim 600$ GeV.
Although the number of events is not large, we may observe the signal since the number of SM background events should be also small for multi-lepton final state.
In the HL-LHC experiments,  we can obtain larger number of events and larger parameter region can be explored.
More detailed simulation study is left as a future work.


\section*{Acknowledgments}
\vspace{0.5cm}
This research is supported by the Ministry of Science, ICT \& Future Planning of Korea, the Pohang City Government, and the Gyeongsangbuk-do Provincial Government (H. O.).
H. O. is sincerely grateful for the KIAS member and all around.


\end{document}